\documentclass[aip,jcp,reprint,floatfix]{revtex4-2}
\usepackage{gensymb}%
\usepackage{verbatim}%
\usepackage{amsmath,amsfonts} %
\usepackage[acronym]{glossaries} %
\usepackage{graphicx} %
\usepackage{dcolumn} %
\usepackage{bm} %
\usepackage{ifpdf} %
\usepackage{color} %
\usepackage{nicefrac} %
\usepackage{natbib} %
\usepackage[breaklinks, %
colorlinks, %
linkcolor=blue, %
citecolor=blue, %
urlcolor=blue]{hyperref} %
\usepackage[table]{xcolor} %

\usepackage[caption=false]{subfig}

\usepackage{braket} %
\usepackage{parskip} %
\newcommand{\bea}{\begin{eqnarray}}
\newcommand{\eea}{\end{eqnarray}}

\def\be{\begin{equation}}
\def\ee{\end{equation}}
\def\bea{\begin{eqnarray}}
\def\eea{\end{eqnarray}}
\def\bal#1\eal{\begin{align*}#1\end{align*}}
\def\ba#1\ea{\begin{align}#1\end{align}}

\begin{document}
                           
\sloppy
\title{
Synthesis technique and electron beam damage study of nanometer-thin single-crystalline Thymine
}

\author{Hazem Daoud$^*$}

\altaffiliation{* These authors contributed equally to this work}
\affiliation{Department of Physics, University of Toronto, Toronto, Ontario, M5S 1A7, Canada}
\author{Sreelaja Pulleri Vadhyar$^*$}
\affiliation{Department of Physics, University of Toronto, Toronto, Ontario, M5S 1A7, Canada}
\affiliation{Department of Chemistry, University of Toronto, Toronto, Ontario, M5S 3H6, Canada}
\author{Ehsan Nikbin$^*$}
\affiliation{Department of Materials Science and Engineering, University of Toronto, Toronto, Ontario, M5S 1A7, Canada}
\author{Cheng Lu}
\author{R. J. Dwayne Miller}
\email {dmiller@lphys2.chem.utoronto.ca}
\affiliation{Department of Physics, University of Toronto, Toronto, Ontario, M5S 1A7, Canada}
\affiliation{Department of Chemistry, University of Toronto, Toronto, Ontario, M5S 3H6, Canada}

\date{\today}

\begin{abstract}
Samples suitable for electron diffraction studies must satisfy certain characteristics such as having a thickness in the range of 10 - 100 nm. We report, to our knowledge, the first successful synthesis technique of nanometer-thin sheets of single-crystalline thymine suitable for electron diffraction and spectroscopy studies. This development provides a well defined system to explore issues related to UV photochemistry of DNA and high intrinsic stability essential to maintaining integrity of genetic information. The crystals are grown using the evaporation technique and the nanometer-thin sheets are obtained via microtoming. The sample is characterized via x-ray diffraction (XRD) and is subsequently studied using electron diffraction via a transmission electron microscope (TEM). Thymine is found to be more radiation resistant than similar molecular moieties (e.g., carbamazepine) by a factor of 5. This raises interesting questions about the role of the fast relaxation processes of electron scattering-induced excited states, extending the concept of radiation hardening beyond photoexcited states. The high stability of thymine in particular opens the door for further studies of these ultrafast relaxation processes giving rise to the high stability of DNA to UV radiation.

\end{abstract}
\maketitle
\glsresetall

\section{Introduction}
Thymine is one of the four molecular building blocks comprising human DNA. It is thus of major interest to study its photochemical dynamics, owing to the importance of mutagenic and carcinogenic effects caused by UV radiation on DNA.\cite{barbatti_aquino_szymczak_nachtigallova_hobza_lischka_2010} The structure of a DNA molecule is stabilized with two hydrogen bonds between thymine and another nucleobase adenine. Thymine plays a vital role in the storage and transfer of genetic features to build living cells. An understanding of its structural relationship and dynamics in conserving genetic information will shed light on the mechanisms of the biological functions of DNA.\cite{bdikin_heredia_neumayer_bystrov_gracio_rodriguez_kholkin_2015} Thymine is particularly important in this regard as it is typically T-T neighboring dimers that undergo photocyclization and is the weak point in DNA with respect to photostability. The quantum yield for this process is less than 5\% but still a significant factor. \cite{schreier_schrader_koller_gilch_crespo-hernandez_swaminathan_carell_zinth_kohler_2007} Understanding the structural dynamics responsible for this degree of photochemical response is important as there is  likely some similar mechanism at play for the other bases that makes them more immune to photochemical disruption in the genetic sequence. Hence, due to its significance there have been many x-ray diffraction, gas phase electron diffraction and spectroscopy studies on thymine.\cite{dementjev_rutkauskas_polovy_macernis_abramavicius_valkunas_dovbeshko_2020, vogt_khaikin_grikina_rykov_vogt_2008, nagpal_dhankhar_cesario_li_chen_rentzepis_2021} Studies on solid thymine are mainly limited to second harmonic generation, coherent anti-Stokes Raman scattering, and electron energy loss spectroscopy. \cite{dementjev2020characterization, isaacson1972interaction1, isaacson1972interaction2} However, to our knowledge, no time-resolved studies of single-crystalline thymine, which would simulate close contact of T-T pairs in DNA, have been carried out yet. Here, there are important questions regarding the ultrafast non-radiative relaxation of excited states that have been determined to occur through a conical intersection (CI).\cite{perun_sobolewski_domcke_2006} The occurrence of CI's in this class of planar aromatic molecules is unique to DNA nucleobases. It has been speculated that this ultrafast non-radiative relaxation and violation of the "energy gap" law conveys inherent "radiation hardening" to UV critical to conserving the genetic code.\cite{barbatti_aquino_szymczak_nachtigallova_hobza_lischka_2010, gustavsson_improta_markovitsi_2010} Moreover, the intersystem crossing and excited state dynamics of cyclobutane pyrimidine dimer (CPD) formed from photo-dimerization of thymines\cite{schreier_schrader_koller_gilch_crespo-hernandez_swaminathan_carell_zinth_kohler_2007} in DNA is of great importance since the formation of photolesions leads to the damage of DNA and interrupts normal cellular processes.\cite{kwok_ma_phillips_2008, mai_richter_marquetand_gonzlez_2017} Thus, thymine represents an important system for UED studies and making single crystals is critical to obtaining the highest possible spatial resolution to directly observe the role of the CI in this key non-radiative relaxation process.\cite{daoud_joubertdoriol_izmaylov_dwaynemiller_2018} Static structures are essential starting points.\newline
In this regard, high-resolution transmission electron microscopy (HRTEM) is an efficient technique for crystal structure determination, the requirement of a sufficiently thin sample sets limitation for this approach. The process of HRTEM imaging consists of two sequential stages. The first step involves interaction of high energy electrons with the sample, producing complex exit wave at the exit surface of the sample. The next step is the navigation of these diffracted waves through the electron optical system where they interact with each other creating recordable images. An increase in specimen thickness, not only reduces the transmission beam but also increases the degree of dynamic scattering of the diffracted beam leading to significant loss of information in image reconstruction. The sample thickness thus plays a crucial role in electron multi-scattering taking place during electron transmission inside the specimen and in exit waves resulting in image formation. \cite{li_chang_wang_xu_ge_2020} \newline
Hence, to achieve the goal of studying this process via electron diffraction, the studied samples must satisfy certain thickness requirements. Preparation of single-crystalline nanometer-thin sheets of thymine for electron diffraction studies has proven particularly difficult owing mainly the the water-solubility of thymine, which renders traditional microtoming ineffective for sample preparation. Cryomicrotome methods could be used. The issues of photostability require studies under ambient conditions relevant to DNA functions. Cryo condition will freeze out potentially important structural relaxation processes. For the most part, studies of thymine via electron diffraction have typically been limited to gas-phase studies.\cite{vogt_khaikin_grikina_rykov_vogt_2008}\newline
 
For comparison with thymine we also conduct an electron beam damage study of carbamazepine of the same thickness and under the same conditions. Carbamazepine is a biomolecule commonly used as an anticonvulsant and pain-relieving drug and is similar in structure to thymine.\cite{leijon1989central} It has been previously studied via electron diffraction under cryo-conditions.\cite{jones2018cryoem}\newline
The purpose of this paper is twofold: (1) We report a successful method for creating nanometer-thin sheets of single-crystalline thymine. (2) We study the electron damage threshold of the sample paving the way to ultrafast electron diffraction studies as well as other pump-probe x-ray diffraction and spectroscopy studies. Although the radiation damage of x-irradiated DNA base pairs has been studied using electron paramagnetic resonance, electron nuclear double resonance, and electron spin resonance spectroscopy \cite{sagstuen1996radiation, huttermann1970electron}, no radiation damage study using electron diffraction on single-crystalline thymine has been carried out.\newline
Studying well defined single crystals could allow for a better understanding of the much debated excited state structural dynamics  of thymine involving inter-system crossing at the atomic level without solvent fluctuations.\cite{ischenko_weber_miller_2017}

\begin{figure*}
\subfloat[Molecular structure.]{
	\centering
		\includegraphics[angle = -90, origin = c, width=0.48 \linewidth]{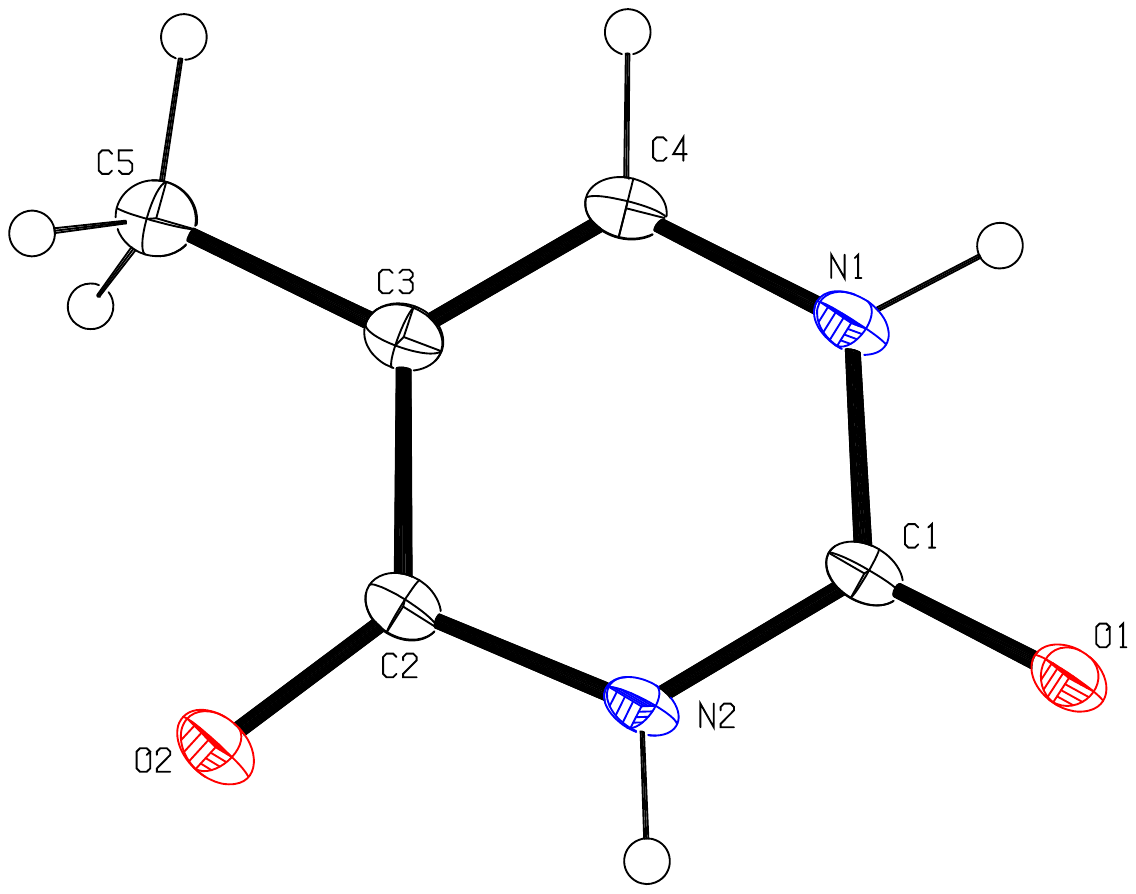}
\label{xrd_diffraction1}
}
\hfill
\subfloat[Crystal unit cell structure.]{
	\centering
		\includegraphics[angle = -90, origin = c, width=0.48 \linewidth]{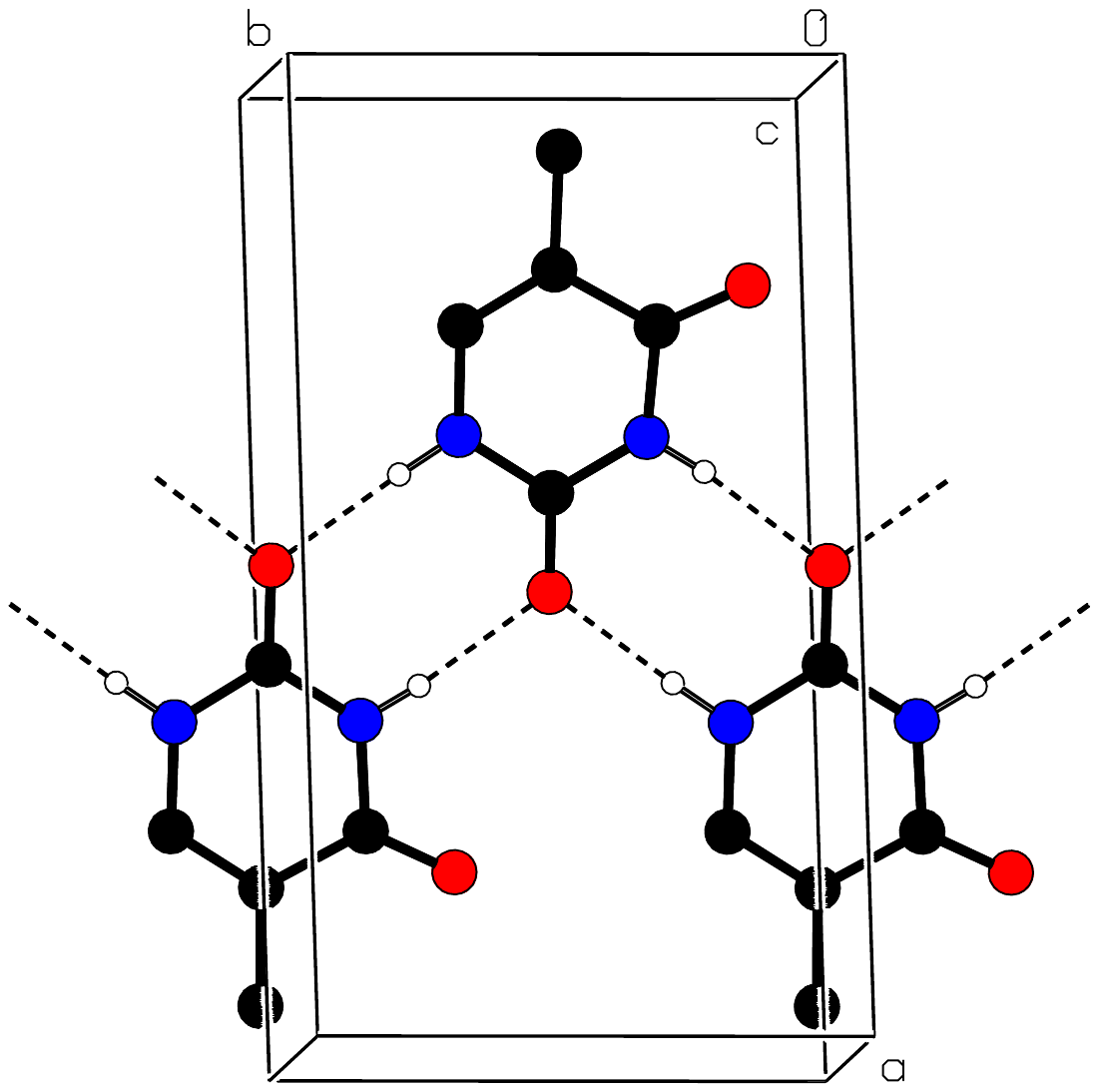}
\label{xrd_diffraction2}
}
	
\caption{Molecular and crystal structure of single crystalline thymine were determined by x-ray diffraction analysis. The study revealed that the grown crystal belongs to monoclinic system, with P2$_{1}$/c space group. The cell parameters are a=12.72 \AA, b= 6.84 \AA, c= 6.62 \AA, $\beta$=104.35\degree and V= 557.9 \AA$^{3}$, Z= 4, which is in close agreement with the reported values.\cite{furberg_hordvik_taugbl_theorell_thorell_1956, ozeki_sakabe_tanaka_1969}}
\label{xrd_results}
\end{figure*}

\begin{table*}
\caption{Crystal data and structure refinement.} \label{table1}
  \centering
  \begin{ruledtabular}
\begin{tabular} {ccc}
 
 Empirical formula & C5 H6 N2 O2    \\
 \hline
Temperature & 150(2) K    \\
\hline
Wavelength & 0.71073 \AA \\
\hline
Crystal system & Monoclinic \\
\hline
Space group & P2$_{1}$/c \\
\hline
Unit cell dimensions & a = 12.722(3) \AA  & $\alpha$ = 90$^{o}$ \\
 & b = 6.8380(14) \AA & $\beta$ = 104.354(7)$^{o}$\\
 & c = 6.6202(13) \AA & $\gamma$ = 90$^{o}$ \\
 \hline
 Volume & 557.9(2) \AA$^{3}$ \\
 \hline
 Z & 4 \\
 \hline
 Density (calculated) & 1.501 Mg/m$^{3}$ \\
 \hline
 Absorption coefficient & 0.119 mm$^{-1}$ \\
 \hline
 Crystal size & 0.720 $\times$ 0.420 $\times$ 0.040 mm$^{3}$

\end{tabular}
  \end{ruledtabular}
\end{table*}

\section{Preparation of single-crystalline nanometer-thin sheets of thymine}
In this section we describe the full procedure for preparing single-crystalline thymine and carbamazepine samples.
\subsection{Preparation of thymine crystals}
Thymine compound purchased from Sigma Aldrich was dissolved in water and the saturated solution was allowed to slowly evaporate at room temperature. Well developed thymine crystals obtained after three weeks were confirmed by single crystal x-ray diffraction (XRD) analysis. Fig. \ref{xrd_results} and table \ref{table1} show the results of the XRD analysis. The study revealed that the grown crystal system is monoclinic, with a P2$_{1}$/c space group. The cell parameters are a=12.72 \AA, b= 6.84 \AA, c= 6.62 \AA, $\beta$=104.35\degree and V= 557.9 \AA$^{3}$, Z= 4, which is in close agreement with the reported values.\cite{furberg_hordvik_taugbl_theorell_thorell_1956, ozeki_sakabe_tanaka_1969}

\subsection{Ultramicrotomy of the crystals}
\begin{figure}
\subfloat[TEM diffraction image of single-crystalline thymine.]{
	\centering
		\includegraphics[width=1 \linewidth]{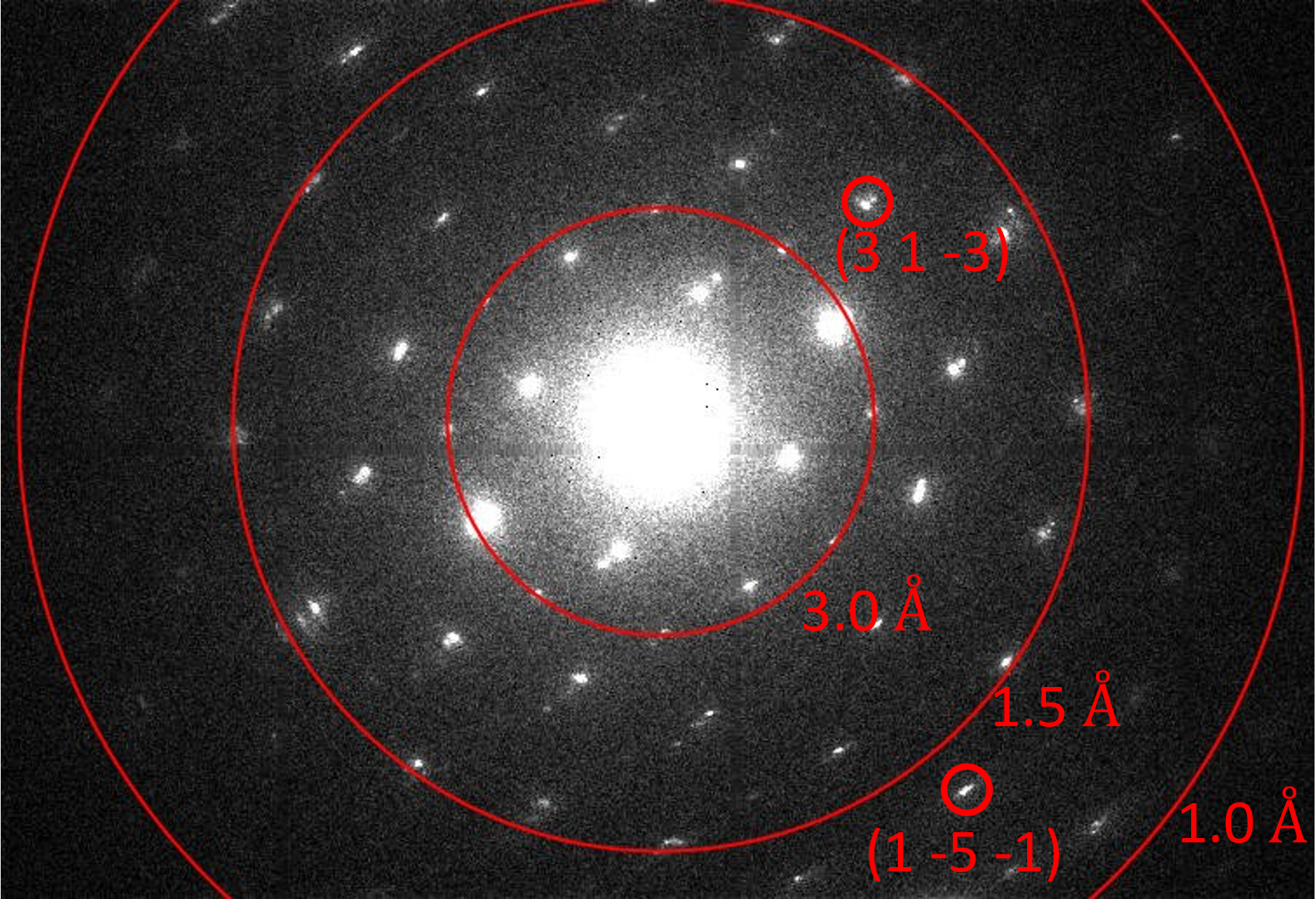}
        
	   \label{thymine_diffraction:sfig1}
    }
\hfill
\subfloat[Intensity decay rates of spots (3 1 -3) and (1 -5 -1) at 2.0 \AA${}$ and 1.3 \AA${}$ resolutions, and characteristic doses of 18 e/\AA$^{2}$ and 13 e/\AA$^{2}$, respectively. The electron dose rate is 0.27 e/(\AA$^{2}$s).]{
	\centering
		\includegraphics[width=1.0 \linewidth]{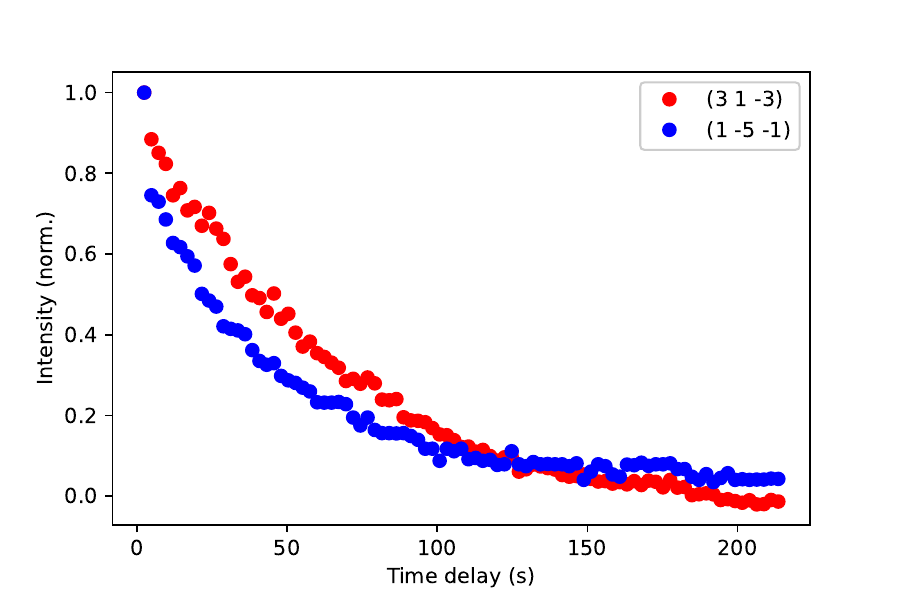}

	   \label{thymine_diffraction:sfig2}
    }
    \caption{Thymine sample electron diffraction data.}
 \end{figure}

Ultramicrotomy is a standard method for ultra thin sample preparation. However, for wet sectioning of water-soluble samples this method is deemed ineffective as the samples dissolve in water before one is able to pick them up from the knife boat and place them on grids. Here, we demonstrate a simple technique to overcome this bottleneck in sample preparation. 
Thymine crystals were carefully embedded into an epoxy resin and cleaved using the Leica Ultracut Ultramicrotome equipped with a PELCO diamond knife. To prevent dissolution of the crystal sections, a saturated aqueous solution of thymine was prepared, and, as opposed to water, we filled the diamond knife boat with the filtered solution in order to collect the sections. Under solution saturation, no crystal dissolution was observed and the uniform thin sections ($\sim$100 nm) remained intact. The sections were carefully picked up and placed over Cu 400 mesh transmission electron microscope (TEM) grids with lacey carbon coating. The initial sections cut were 300 nm thin and they were discarded till perfect flat crystal sections were obtained. Then the thickness setting was reduced in steps of 50 nm. Finally, 100 nm sections were used for the experimental electron damage studies. This sample preparation procedure provides a novel and reliable way of obtaining thin sections for many water-soluble samples from bulk crystals for electron diffraction and spectroscopy studies.\newline Similar to thymine, prismatic crystals of carbamazepine (form III) were grown in ethanol at room temperature. The crystal structure was confirmed via XRD analysis. 100 nm thin sheets were obtained through microtoming. The sections, floated on saturated carbamazepine solution, were captured and analysed using the TEM. The thymine and carbamazepine diffraction patterns were indexed and the zone axis of each crystal was determined to be [1 0 1] and [0 0 1], respectively. The diffraction patterns of thymine and carbamazepine are shown in Figs. \ref{thymine_diffraction:sfig1} and \ref{CBZ1}, respectively.\newline 
The obtained thymine samples were studied via a TEM. The thymine samples proved to be robust under vacuum conditions and lasted for many weeks without any signs of degradation. This observation alone illustrates the robust nature of thymine.\newline


\section{TEM electron damage study}
To study the electron damage of thymine and carbamazepine samples, we monitored the intensities of the diffraction peaks until they dropped to 1/e of their initial values, and until the crystal structures were completely destroyed and no diffraction peaks could be observed. We compared the recorded electron dose rates for both samples. The experiments were repeated multiple times and at various fluences.
 \subsection{Experimental setup and procedure} 
 Single crystal, 100 nm thin, thymine samples were loaded into a Hitachi HT-7700 TEM, which was operated at a voltage of 80 kV, and is equipped with an X-Spectrum Amber 750K direct electron detector. 
 The diffraction patterns were acquired automatically using a custom written Python script to control the electron beam intensity and the detector image acquisition. The control of the microscope and the detector was enabled using the \textit{hihi} package developed by Hitachi High-Tech Inc$.$ Canada. Various beam intensities were achieved by focusing the beam on the sample and the sequences of diffraction patterns over time were acquired. The diffraction spots were identified using a custom script. The total intensities within a radius of 5 pixels were added up and tracked over time.

 \subsection{Analysis}

 \subsubsection{Effect of accumulated electron dose}
\begin{figure}
\subfloat[TEM diffraction image of single-crystalline carbamazepine.]{
	\centering
		\includegraphics[width=1 \linewidth]{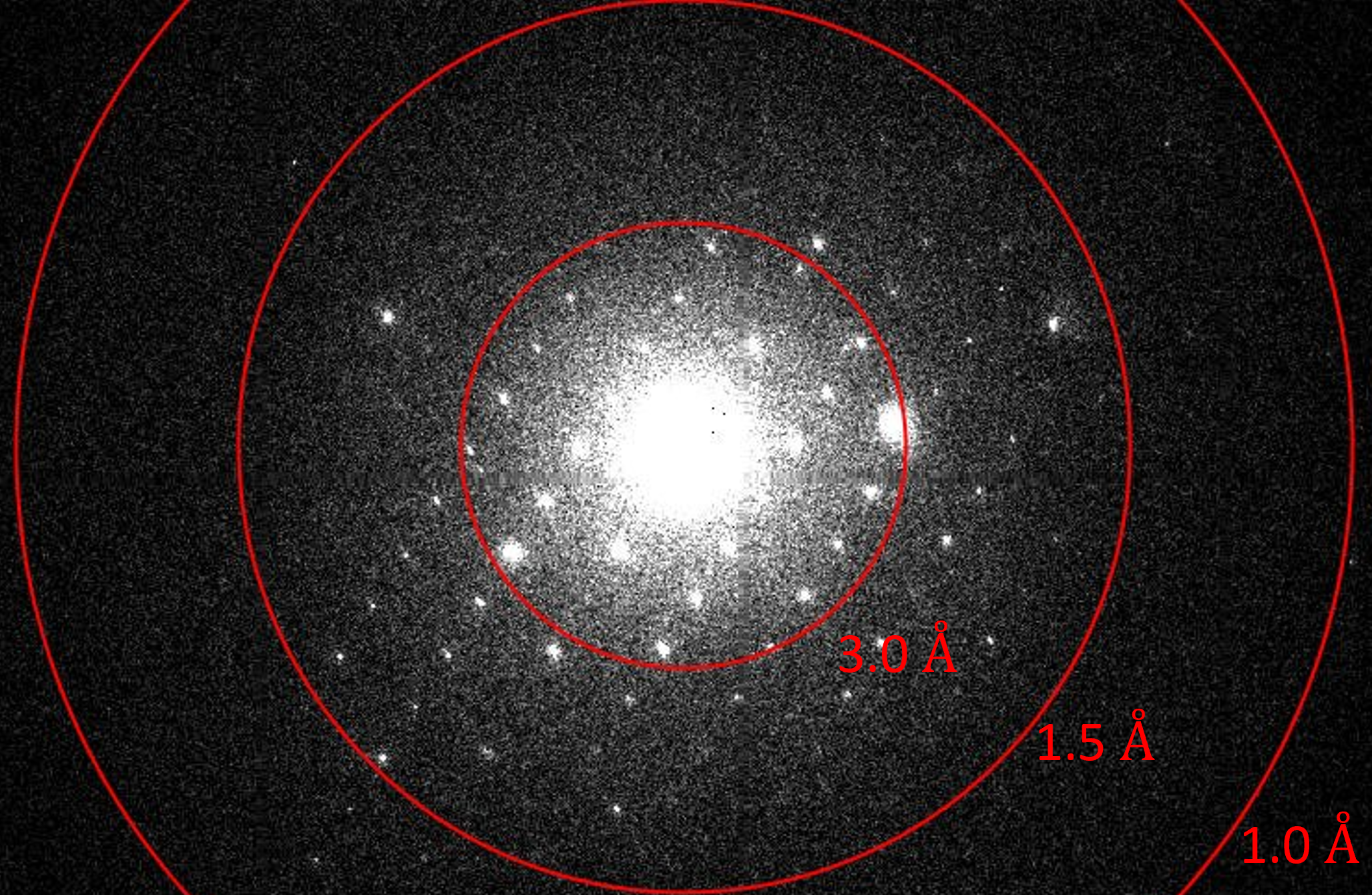}
        
	   \label{CBZ1}
    }
\hfill
\subfloat[TEM diffraction image of single-crystalline carbamazepine after irradiation with a dose of 16 e/\AA$^{2}$, the critical dose resulting in complete destruction of the diffraction pattern. \cite{egerton2004radiation}]{
	\centering
		\includegraphics[width=1 \linewidth]{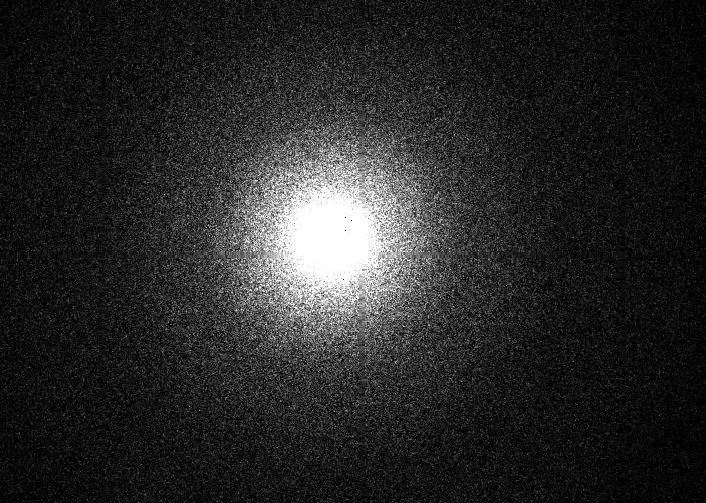} 
	   \label{CBZ2}
    }
    \caption{Carbamazepine diffraction patterns before and after electron irradiation.}
 \end{figure}

Fixing the electron voltage, we monitored the effect of the accumulated electron dose on different diffraction spots in the diffraction pattern. Fig. \ref{thymine_diffraction:sfig2} shows the intensity changes for spots (3 1 -3) and (1 -5 -1) as a function of time for an electron dose rate of 0.27 e/(\AA$^{2}$s) at 80 kV. The graphs show exponential decay in intensity for both spots. The spot (3 1 -3) corresponds to the inter-planar d-spacing of 2.0 \AA, while (1 -5 -1) is a higher resolution spot corresponding to the d-spacing of 1.3 \AA. The higher resolution spot fades away faster. We monitored the exponential decay in intensity and we recorded total electron dose where the intensity drops to 1/e of the initial value, known as the characteristic dose.  \cite{egerton2004radiation, egerton2012mechanisms} The intensity of the (3 -1 -3) spot reduces to 1/e of its initial value at the accumulated electron dose of  $\sim$ 18 e/\AA$^{2}$ compared to $\sim$ 13 e/\AA$^{2}$ for the spot (1 -5 -1). The beam damage sensitivity was tested at the dose rates of 0.54 e/(\AA$^{2}$s) and 0.15 e/(\AA$^{2}$s) and similar results were obtained. \\
We have also determined the total electron dose required for complete destruction of the diffraction pattern of thymine and carbamazepine (Fig. \ref{CBZ2}), known as critical dose. \cite{egerton2004radiation} The critical dose for thymine was determined to be $\sim$ 85 e/\AA$^{2}$ compared to $\sim$ 16 e/\AA$^{2}$ for carbamazepine, suggesting that thymine is around 5 times more radiation resistant than carbamazepine. The obtained critical doses are the average of five different measurements. 
Other studies have shown that solving the structure of small molecules like acetaminophen, biotin, carbamazepine, etc. at $\sim$1 \AA${}$ resolution requires keeping the electron dose as low as 3 e/\AA$^{2}$ at cryogenic temperatures at an electron energy of 200 keV. \cite{jones2018cryoem} The typical electron dose for structure determination of protein nanocrystals at cryogenic temperatures at $\sim$1.7 \AA${}$ resolution is around 1 e/\AA$^{2}$ to 5 e/\AA$^{2}$. \cite{nannenga2019cryo, bucker2020serial} Electron radiation damage, due to knock-on damage, at 200 keV is typically twice the damage at 80 keV for the elements carbon, nitrogen and oxygen.\cite{hobbs1979introduction} Additionally, at cryogenic temperatures the electron beam-induced damage is considerably reduced as compared to room temperature.\cite{knapek1980beam} Taking these points into consideration, we observe that the threshold for electron radiation damage for thymine is considerably higher than similar classes of molecules. This observation could be due to the involvement of very fast relaxation of electron excited upper states in the scattering process via the inverted conical intersection connecting upper electronic states in this interesting molecular system. We see here already intrinsic radiation hardening that is intrinsic to the thymine structure. However, further studies are needed to verify this hypothesis. The predominant damage mechanism for organic materials is radiolysis at the electron energy of 80 keV.  \cite{egerton2012mechanisms} However, knock on damage can also occur due to displacement of hydrogen atoms. \cite{egerton2004radiation} \newline 
We observe that the intensity of the diffraction spot (1 -5 -1) at 1.3 \AA${}$ resolution drops to 1/e its initial value at an approximate dose of 13 e/\AA$^{2}$. In a UED experiment with 10$^{6}$ electrons per pulse at 1 kHz and a beam diameter of 100 $\mu$m the electron dose rate is $\sim$1.3 $\times$ 10$^{-3}$ e/\AA$^{2}s$, so it would take more than 10$^{4}$ s to induce substantial damage on the sample, which is enough time to obtain a strong signal.

 \section{Conclusion}
 We have successfully prepared the first single-crystalline nanometer-thin thymine samples that are suitable for electron diffraction studies. The samples are robust in vacuum and did not show any signs of degradation over a span of several months. The crystal structure and characteristics were determined using XRD. The crystals were studied using electron diffraction via a TEM and the characteristic and the critical electron doses were determined for thymine and carbamazepine. Thymine was found to be extremely robust in terms of stability under vacuum conditions at room temperature. The degree of robustness was further exemplified in the factor of 5 higher threshold with respect to the critical electron dose for radiation damage. This high threshold for electron-induced damage could be related to similar relaxation pathways through conical intersection within electronic surfaces that rapidly converts excited states from either photo-induced or electron scattering-induced excitation. Radiolysis and displacement of hydrogen atoms could also be dominant mechanisms for the electron beam damage of thymine or carbamazepine. Further studies are needed to determine the exact mechanisms involved.\newline
 This study opens the door for conducting time resolved serial diffraction studies to directly observe the atomic motions coupled to excited state relaxation processes that lead to the exceptional UV stability of DNA nucleobases and the prospect to direct imaging of conical intersection processes in this important class of molecules.

 \section{Acknowledgement} 
 The authors acknowledge the Open Center for Characterization of Advanced Materials (OCCAM) facility for the TEM support and Hitachi High-Tech Canada Inc$.$ for providing the tools for automated data acquisition. 
\bibliography{mybib}
\end{document}